\documentclass[twocolumn]{article}
\usepackage{graphicx} %
\usepackage{url}
\usepackage{amsthm,amsmath,amssymb}
\usepackage{authblk}
\newcommand{\ind}{\perp\!\!\!\!\perp}
\usepackage{amsthm,amsmath,amssymb,enumerate}

\usepackage{enumitem}
\usepackage{booktabs}
\usepackage{graphicx}
\usepackage{subfig}
\usepackage{subcaption}
\usepackage{tikz}
\usepackage{dsfont}
\usepackage{graphicx} 
\usepackage{amsmath, amssymb, amsthm} 
\usepackage{newtxtext, newtxmath}

\usepackage{natbib}
\providecommand{\keywords}[1]{\par\addvspace\baselineskip
\noindent\small\textbf{\textit{Keywords---}}\ignorespaces#1}

\theoremstyle{remark}
\newtheorem{rem}{Remark}

\title{Sensitivity Analysis of the Consistency Assumption}
\author[1]{Brian Knaeble\thanks{email: bknaeble@uvu.edu}}
\author[2]{Qinyun Lin}
\author[3]{Erich Kummerfeld}
\author[4]{Kenneth A. Frank}

\affil[1]{Department of Computer Science, Utah Valley University}
\affil[2]{School of Public Health and Community Medicine, University of Gothenburg}
\affil[3]{Institute for Health Informatics, University of Minnesota}
\affil[4]{Department of Counseling, Educational Psychology, and Special Education, Michigan State University}

\begin{document}
%
\maketitle
\begin{abstract}
Sensitivity analysis informs causal inference by assessing the sensitivity of conclusions to departures from assumptions. The consistency assumption states that there are no hidden versions of treatment and that the outcome arising naturally equals the outcome arising from intervention. When reasoning about the possibility of consistency violations, it can be helpful to distinguish between covariates and versions of treatment. In the context of surgery, for example, genomic variables are covariates and the skill of a particular surgeon is a version of treatment. There may be hidden versions of treatment, and this paper addresses that concern with a new kind of sensitivity analysis. Whereas many methods for sensitivity analysis are focused on confounding by unmeasured covariates, the methodology of this paper is focused on confounding by hidden versions of treatment. In this paper, new mathematical notation is introduced to support the novel method, and example applications are described.
\end{abstract}
\section{Introduction}
%
To enable the drawing of causal conclusions from data, the Rubin Causal Model assumes consistency as part of the stable unit treatment value assumption (SUTVA) \cite[Section 1.6]{Imbens2015}. Consistency is distinct from an assumption of no interference, and sometimes considered axiomatic \cite{Pearl2010b}. Consistency means that there are no hidden versions of treatment and that the outcome arising naturally equals the outcome arising from intervention \cite{VanderWeele2009}. While there are methods for conducting causal inference under multiple versions of treatment \cite{VanderWeele2013,Hasegawa2020}, there are good reasons to suspect that consistency is often violated in practice \cite{Rehkopf2016,Weichenthal2022}.


The problem of inconsistency can be addressed with sensitivity analysis, similar to how the problem of confounding has been addressed \cite[Chapter 3]{Knaeble2023,SAWA,Rose10}. This paper describes the results of a mathematical investigation that has yielded
\renewcommand{\theenumi}{\roman{enumi})}%
\begin{enumerate}
    \item a formal framework within which to analyze violations of the consistency assumption,
    \item a novel sensitivity parameter to represent bias induced by consistency violations, and
    \item methods for specifying bounds on components of that sensitivity parameter to facilitate partial identification of causal estimands.
\end{enumerate} 
When interpreting the results of causal inference studies, whether observational or experimental, consideration of the value of the novel sensitivity parameter helps to clarify the size and direction of bias due to violations of consistency.

To prepare to describe the methodology and its application in the context of an observational study, and then in the context of an open-label clinical trial, an introductory example is described here in a familiar context to help improve intuition for the mathematical notation, definitions, and results that follow.
%
\subsection{A playful example}
\label{golf}
\begin{figure}[ht]
\centering
\includegraphics[width=1.0\linewidth]{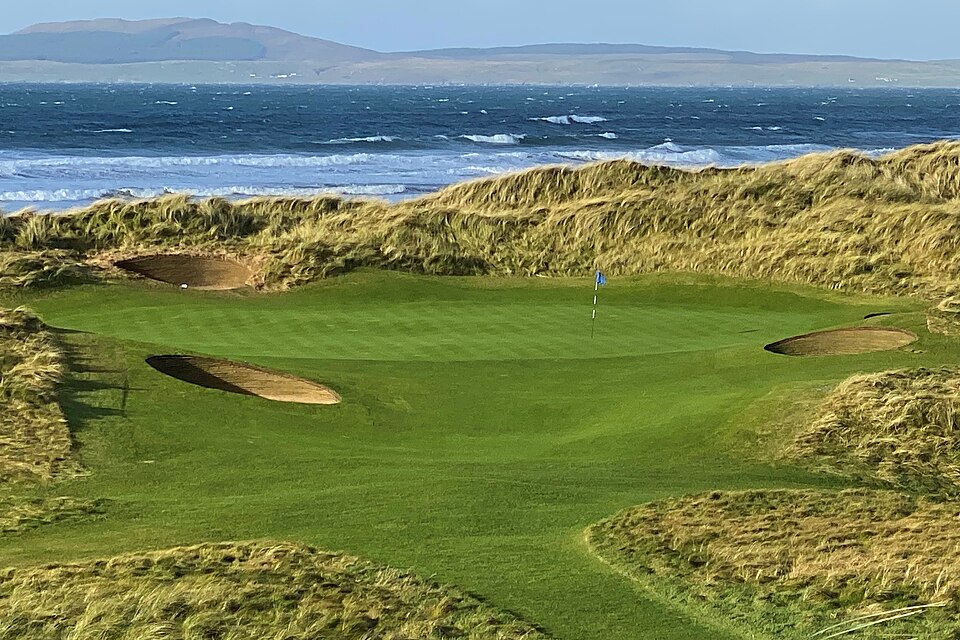}
\caption{The par 3, 9th hole at The Machrie Links in Scotland \cite{Deanamuir}.}
\label{fig1}
\end{figure}

%
The data set in Table \ref{tab1} shows relative frequencies that could have been observed by watching many golfers play a par 3 hole. On a par 3 hole, a golfer hits first from a tee box, and it can be difficult to forecast whether or not the tee shot will land on the putting green, especially in windy conditions. An uncertain approach to a putting green with various hazards is shown in Figure \ref{fig1}. On a par 3 hole, the tee shot is expected to land near the green, but when the ball comes to rest after that first shot, its position may or may not be on the putting green.

In this hypothetical scenario, the treatment variable is the type of club (putter or iron) used on the second shot. Typically, a golfer will putt from the putting green and chip (with an iron) from the fringe or rough surrounding the putting green, although golfers occasionally choose to putt from off the green. The outcome variable is the eventual score on the hole ($\leq 2$, or $\geq 3$). For simplicity, the successful outcome ($\leq 2$) is referred to here as a birdie even though a rare hole-in-one (an eagle, on a par 3) is possible. Do putts cause birdies?  
\begin{table}[ht!]
\centering
\caption{A contingency table of synthetic data with relative frequencies and a positive association (relative risk, $RR=10$) between putting on the second shot and a low score (a birdie or an eagle) on a par 3.}
\label{tab1}
\begin{tabular}{rcc}
\toprule
 & \multicolumn{2}{c}{Club on 2nd shot}\\
\cmidrule{2-3}
Score&Iron&Putter\\
\hline
$\leq 2$&0.5\%&5\%\\
$\geq 3$&49.5\%&45\%\\
\hline
\end{tabular}
\end{table}

The association of Table \ref{tab1} could be entirely spurious. One non-causal explanation appeals to a heterogeneous population of golfers. On the tee box before any shots, the better golfers have higher propensity probabilities for putting on the second shot and better prognoses for birdie. Another non-causal explanation appeals to different versions of treatment. The position of the ball after the first shot matters. Here, position is interpreted as a version of treatment. 

Suppose that each golfer putts if and only if the ball is on the green. For each golfer, the position of the ball after the first shot can be modeled as a random variable. Putting is thus (quasi) randomly ``assigned,'' but the position of the ball after the first shot can be conceptualized as both a version of treatment and an invalid instrument that violates the exclusion restriction \cite{Baiocchi2014}. 

Conditioning on position is an option, but it eliminates much randomness from the natural process that leads to treatment \cite{Knaeble2023}. It can be problematic to condition on the results of a noisy process \cite{deming1994new}, and conditioning on an instrument can lead to bias amplification \cite{Wool,Pim}. However, it is also problematic to ignore a severely imbalanced covariate like distance \cite{Rubin2009}. These problems can be addressed with sensitivity analysis, and the sensitivity analysis can also handle hidden versions of treatment.
%

Although ball position is an obvious version of treatment, it is only one of several possible versions of treatment that arise naturally in this setting. How the ball lies in the grass is an example of a moderately hidden version of treatment. Various anxieties in the mental state of the golfer are fully hidden versions of treatment. These kinds of complications, in which unrecorded versions directly influence the outcome, have been documented in applied settings \cite{Rehkopf2016,Mellon2020}, thus warranting further study.
\subsection{Indeterminism}
Instrumental variables techniques can be used to estimate the local average treatment effect on a subpopulation of compliers \cite{angrist1994ident}. This was done famously within the context of a natural experiment, with a lottery as the instrument \cite{angrist1990lifetime}. However, the subpopulation of compliers is defined via principal strata \cite{Frangakis2002}, reflecting an assumption of Laplacian determinism \cite{Dawid2012}. This means that each individual is assumed to have fixed potential behaviors, such as whether they would comply or not, under encouragement. 

Notation can sometimes hide assumptions \cite[p. 881]{VanderWeele2009}, and the consistency assumption is hidden in standard potential outcomes notation, although it has been recognized as an assumption \cite{Cole2009}. When there are hidden versions of treatment, stochastic potential outcomes are recommended \cite{VanderWeele2009}. Statistical modeling with stochastic potential outcomes has received some attention recently \cite{Gelman2025}. Building on that momentum, we formulate a framework for sensitivity analysis of the consistency assumption with stochastic potential outcomes. We use mathematical notation with proven utility for the purpose of studying stochastic processes. 
\section{Notation and definitions}
\label{nd}
The population size $N$ must be large enough so that the equality constraints of (\ref{probs}) are approximately satisfied by the law of large numbers. Also, $N$ should be relatively large compared to the size of the space of versions and the sizes of the probability spaces. The individuals in the population are assumed to be jointly independent and indexed by $i$ in $I:=\{1,...,N\}$. 

\subsection{Probability spaces}
A filtered probability space is a probability space with a filtration, and a filtration is a sequence of nested $\sigma$-algebras \cite[p. 458]{Billing} representing increasing information over time. 

Each individual $i$ has a natural, filtered probaiblity space associated with the data generating process. It is denoted by \[(\Omega_i,\mathcal{F}_{i,2},\mathbb{F}_{i},P_{i}),\] where $\Omega_{i}$ is a sample space and $P_{i}$ is a probability measure. Its filtration $\mathbb{F}_{i}=(\mathcal{F}_{i,0},\mathcal{F}_{i,1},\mathcal{F}_{i,2})$ satisfies \[\mathcal{F}_{i,0}\subseteq \mathcal{F}_{i,1}\subseteq \mathcal{F}_{i,2}.\]

The sub $\sigma$-algebra $\mathcal{F}_{i,0}$ represents background information available at the outset of a trial, before a quasi-random process that leads to treatment plays out. The sub $\sigma$-algebra $\mathcal{F}_{i,1}$ represents information available at the moment when treatment is determined. The $\sigma$-algebra $\mathcal{F}_{i,2}$ represents information at the end of the trial when the outcome is measured. The $\sigma$-algebras consist of theoretically measurable but latent sets.
\begin{rem}
    In the golf context of Section \ref{golf}, the measurable sets in $\mathcal{F}_{i,0}$ are what is theoretically measurable on the tee box, the measurable sets in $\mathcal{F}_{i,1}$ are what is theoretically measurable  after the first shot and before the second shot, and the measurable sets in $\mathcal{F}_{i,2}$ are what is theoretically measurable after the second shot.
\end{rem}

Each individual $i$ has two, interventional, probability spaces
\[\{(\Omega_{i,x},\mathcal{F}_{i,x,2},\mathbb{F}_{i,x},P_{i,x})\}_{x\in\{0,1\}}.\]
The subscript $x$ signals whether treatment has been set by a hypothetical intervention \cite{VanderWeele2018} to one of two levels. As with the natural probability space, here in the interventional setting, for each $x\in\{0,1\}$, $\Omega_{i,x}$ is a sample space, $P_{i,x}$ is a probability measure, and the filtration $\mathbb{F}_{i,x}=(\mathcal{F}_{i,x,0},\mathcal{F}_{i,x,1},\mathcal{F}_{i,x,2})$ satisfies \[\mathcal{F}_{i,x,0}\subseteq \mathcal{F}_{i,x,1}\subseteq \mathcal{F}_{i,x,2}.\]
Prior to each intervention,
\begin{align}
   \nonumber&\forall i\in I,\\ &\nonumber(\Omega_i,\mathcal{F}_{i,0},P_i)\cong(\Omega_{i,0},\mathcal{F}_{i,0,0},P_{i,0})\cong(\Omega_{i,1},\mathcal{F}_{i,1,0},P_{i,1}),
\end{align}
where $\cong$ means isomorphic. We will sometimes assume
\begin{align}
   \label{first}&\forall i\in I,\\ &\nonumber(\Omega_i,\mathcal{F}_{i,1},P_i)\cong(\Omega_{i,0},\mathcal{F}_{i,0,1},P_{i,0})\cong(\Omega_{i,1},\mathcal{F}_{i,1,1},P_{i,1}).
\end{align}
%

Additionally, there is a population, probability space $(I,2^I,\textrm{Pr})$. Its sample space is $I$, all of its subsets are measurable, and the probability measure $\textrm{Pr}$ measures the proportion of individuals with an $\mathcal{F}_{i,0}$-measurable attribute.

\subsection{Random variables}
A random variable is a measurable function \cite[p. 182]{Billing}. 

The individual treatment random variable $X_i$ has the following properties: it is a function on $\Omega_i$, it is $\mathcal{F}_{i,1}$-measurable, and it takes the value $x\in \{0,1\}$ with probability $P_i(X_i^{-1}(x))$. For each $x\in\{0,1\}$, $X_i$ is well-defined on $\Omega_{i,x}$ and $\mathcal{F}_{i,x,1}$-measurable, under (\ref{first}).

The individual outcome random variable $Y_i$ has the following properties: it is a function on $\Omega_i$, it is $\mathcal{F}_{i,2}$-measurable, and it takes the value $y\in \{0,1\}$ with probability $P_i(Y_i^{-1}(y))$.

\begin{rem}
    Here, for simplicity and concreteness, we consider only treatments and outcomes that are dichotomous categorical, but some
    of the ideas that follow can be generalized for application with general categorical or continuous variables.  
\end{rem}

For each $x\in\{0,1\}$ the stochastic potential outcome
\begin{equation}\label{stochdef}Y_{i,x}\end{equation}
is a $\mathcal{F}_{i,x,2}$-measurable random variable on $\Omega_{i,x}$ that takes the value $y\in \{0,1\}$ with probability $P_{i,x}(Y_{i,x}^{-1}(y))$.
\subsection{Expectations}
The following expectations are $\mathcal{F}_{i,0}$-measurable attributes. Define the propensity probabilities $\{\pi_i:=P_i(X_i=1)\}_{i\in I}$, the interventional prognosis probabilities
\begin{equation}\label{ints}\{r_i(x):=E(Y_{i,x}\}_{i\in I,x\in\{0,1\}},\end{equation} and the natural, conditional prognosis probabilities \begin{equation}\label{nats}\{r_{i|x}:=E(Y_i|X_i=x)\}_{i\in I,x\in\{0,1\}}.\end{equation}

Those natural, conditional, prognosis probabilities are expectations of the following $\mathcal{F}_{i,1}$-measurable random variables: $\{R_{i|x}:=E(Y_i|\mathcal{F}_{i,1})|(X_i=x)\}_{x\in\{0,1\},i\in I}$.

\begin{rem}
    To understand the meaning of each $R_{i|x}$ it may help to revisit the golf context of Section \ref{golf}, where $R_{i|1}$ represents, for individual $i$, the distribution of prognosis probabilities for birdie, depending on how the first shot goes, conditional on the requirement that the first shot naturally leads to putting. The meaning $R_{i|0}$ is analogous but with chipping instead of putting.
\end{rem}

\subsection{Observations}
For each $i\in I$, a trial is conducted to observe the $\mathcal{F}_{i,1}$-measurable treatment value $X_i\in\{0,1\}$ and the $\mathcal{F}_{i,2}$-measurable outcome value $Y_i\in \{0,1\}$. All those trials result in population-wide, observational data $\{x_i,y_i\}_{i\in I}$.

\subsection{Related assumptions}
Define consistency as
\begin{equation}\label{0consdef}\forall i\in I,x\in\{0,1\}, (Y_i|(X_i=x))=^d Y_{i,x},\end{equation}
where $=^d$ means equality in distribution.
When $Y$ is dichotomous categorical then consistency as defined in (\ref{0consdef}) is equivalent to
\begin{equation}\label{0conspar}
\forall i\in I,x\in\{0,1\}, r_{i|x}=r_i(x).
\end{equation}

The $6$-tuple \[(\pi_i,x_i,r_{i|0},r_{i|1},r_i(0),r_i(1))\] is a random variable over the sample space $I$. Denote that joint random variable with \[(\dot\pi,\dot{x},\dot r_0,\dot r_1,\dot r(0),\dot r(1)).\] 

With $\ind_I$ denoting approximate independence over $I$, we define background unconfounding as 
\begin{equation}\label{back1} (\dot r_0,\dot r_1)\ind_I \dot \pi.\end{equation}

With $\ind$ denoting approximate independence over any $\Omega_i$, we define version unconfounding as
\begin{equation}
\label{longconf}
\forall i\in I,(R_{i|0},R_{i|1})\ind X_i.
\end{equation}

If (\ref{back1}) and (\ref{longconf}) are true, then by the law of large numbers, we have
\begin{equation}\label{back2} (\dot r_0,\dot r_1)\ind_I \dot x.\end{equation}
Under consistency as defined in (\ref{0consdef}), the independence condition in (\ref{back2}) is by (\ref{0conspar}) equivalent to
\begin{equation}\label{0back3} (\dot r(0),\dot r(1))\ind_I \dot x.\end{equation}

We do not accept any of the assumptions in (\ref{0consdef}-\ref{0back3}). Those assumptions have been stated to clarify the meaning of concepts related to our methodology. Unconfoundedness as defined in (\ref{0back3}) is similar to the definition put forth by Rosenbaum \cite[Section 3.3]{Rose10}. It follows from the definition put forth by Imbens and Rubin \cite[Definition 3.6]{Imbens2015} by the law of large numbers. It could be referred to as an assumption of strongly ignorable treatment assignment, if we assume also that treatment assignment is probabilistic, meaning $\forall i\in I, 0<P_i(X_i=1)<1$. The stochastic consistency definition in (\ref{0consdef}) reflects conditions, concerns, and discussion of consistency by VanderWeele \cite{VanderWeele2009}. The definition in (\ref{0consdef}) captures what VanderWeele refers to as the real meaning of the consistency assumption \cite[Condition C2]{VanderWeele2009}. VanderWeele distinguishes between consistency and treatment-variation irrelevance \cite[Condition C1]{VanderWeele2009}. There can be relevant versions of treatment that do not lead to bias if (\ref{longconf}) holds, and there can be relevant background characteristics that do not lead to bias if (\ref{back1}) holds. In the two-staged sensitivity analysis that follows, we relax both (\ref{back1}) and (\ref{longconf}). That methodology is described in Section \ref{methods}.
%
%
\subsection{Estimands}
Based on the definitions in (\ref{ints}) we define the causal estimand
\begin{equation}
    \label{taucausal}
    \tau:=\sum_{i=1}^N (r_i(1)-r_i(0))/N.
\end{equation}
Based on the definitions in (\ref{nats}) we define a related (but not necessarily causal) estimand
\begin{equation}
    \label{psinoncausal}
    \psi:=\sum_{i=1}^N (r_{i|1}-r_{i|0})/N.
\end{equation}
\begin{rem}
    The methods of this paper are based on differences between prognosis probabilities, but adaptation to other measures is possible.
\end{rem}
%
%
\section{Mathematical methods}
\label{methods}
This Section splits the mathematics of sensitivity analysis into two distinct analyses. The first analysis relaxes the assumption of background unconfounding as defined in (\ref{back1}). The second analysis relaxes the assumption of version unconfounding as defined in (\ref{longconf}). 
\subsection{Analysis of background confounding}
\label{backsection}
%
Denote the unit, 3-cube with $C=\{(\pi,r_{0},r_{1}):0<\pi,r_{0},r_{1}<1\}$, where each of $\{\pi,r_0,r_1\}$ is now a mathematical variable. 
Define $r=\pi r_1+(1-\pi)r_0$.
The space of distributions on $C$ is denoted with $\mathcal{P}(C)$, and we write $\mu$ for a generic distribution in $\mathcal{P}(C)$. The observed data $\{(x,y)_i\}_{i\in I}$ is now thought of as a joint random variable $(x,y)$ over $I$.

We may partially identify $\psi$ by solving the following constrained optimization problems with linear programming.
\begin{subequations}
\label{probs}
\begin{align}
\label{objective}
&\psi_{\textrm{min/max}}  \  \approx \ \\
\label{integrand}&\underset{\mu\in\mathcal P(C)}{\textrm{min/max}}\ \int_C (r_1-r_0) d\mu
, \\ &\textrm{subject to } \ \\& 
\int_C (1-\pi)r_0 d\mu \ = \ \textrm{Pr}(x=0,y=1), \\
\label{ctwo}
& \int_C \pi r_1 d\mu \ = \ \textrm{Pr}(x=1,y=1), \\
\label{cthree}
& \int_C (1-\pi)(1-r_0)d\mu \ = \ \textrm{Pr}(x=0,y=0), \\
\label{cfour} 
& \int_C \pi(1-r_1)d\mu  \ = \ \textrm{Pr}(x=1,y=0),\\
\label{ione} 
& \int_C (\pi-\textrm{Pr}(x=1))^2d\mu \leq f,  \textrm{~and} \\
\label{itwo} 
& \int_C (r-\textrm{Pr}(y=1))^2d\mu \leq g.
\end{align}
\end{subequations}

\begin{rem}
    There are alternative techniques to partially identify $\psi$ based on other related constraints. For instance, a lower bound for entropy \cite{KnaEnt} could replace (\ref{ione}) or (\ref{itwo}).
\end{rem}

The equality constraints in (\ref{probs}) are expected to hold in the limit of large $N$ by the law of large numbers. Resampling methods can be applied if the population is of smaller size \cite{Knaeble2024}. Subject matter knowledge determines $f$ and $g$ in lines (\ref{ione}) and (\ref{itwo}). The upper bounds $f$ and $g$ are smaller when there is more randomness \cite{Knaeble2023} or entropy \cite{KnaEnt} in the data generating process. So-called natural, natural experiments are natural experiments that attempt to harness the ``almost perfect randomness'' that nature occasionally provides \cite{Rosen}. We expect a smaller $f$ in a natural experiment, leading to tighter bounds on $\psi$.  Measures of discordance from twin studies can be transported for empirical specification of $f$ and $g$ \cite[Section 2.2, Proposition 3.2, Appendix A]{knaeblePartial}. Alternatively, $f$ and $g$ can be estimated using Tjur's coefficient of discrimination \cite[Theorem 3.3]{Tjur2009,KnaEnt}.

By solving \ref{probs} we obtain $\psi_{\min/\max}$ values which we denote with $L=\psi_{\min}$ and $U=\psi_{\max}$ so that the interval $(L,U)$ satisfies \begin{equation}\label{psibound}L\leq \psi \leq U.\end{equation}
In a large randomized trial $f=0$ and we may identify $L=\psi=U$, but $\tau\neq \psi$ is possible in the absence of adequate experimental control and blinding.
\subsection{Analysis of version confounding}
It remains to conduct sensitivity analysis of version confounding. Toward that end, we define, $\forall i\in I$, \[K_i:=(r_{i|1}-r_i(1))-(r_{i|0}-r_i(0))\] and introduce the following sensitivity parameter, a simple bias parameter:
\begin{equation}
    \label{senspar2}
    K=\sum_{i=1}^N K_i/N.
\end{equation}

Based on the definitions in (\ref{taucausal}) and (\ref{psinoncausal}), combining (\ref{psibound}) and (\ref{senspar2}) results in 
\begin{equation}
    \label{maininterval}
    L-K<\tau<U-K.
\end{equation}

To partially identify $\tau$ we compute $(L,U)$ with a technique of Section \ref{backsection} and then specify $K$ on the basis of subject matter knowledge. 

In general, to specify $K$, we may reason about the values in $\{r_{i|1},r_i(1),r_{i|0},r_i(0)\}_{i\in I}$. 

Under (\ref{first}), each $K_i$ can be decomposed as follows:
\begin{subequations}
\begin{align}
\nonumber K_i&=(r_{i|1}-r_i(1))-(r_{i|0}-r_i(0))\\
\nonumber&=\pi_ir_{i|1}+(1-\pi_i)r_{i|1}-\pi_iE(Y_{i,1}|X_i=1)\\
\nonumber&\quad-(1-\pi_i)E(Y_{i,1}|X_i=0)-\pi_ir_{i|0}-(1-\pi_i)r_{i|0}\\
\nonumber&\quad+\pi_i E(Y_{i,0}|X_i=1)+(1-\pi_i)E(Y_{i,0}|X_i=0)\\
\label{AA}&=\pi_i(r_{i|1}-E(Y_{i,1}|X_i=1))\\
\label{BB}&\quad-(1-\pi_i)(r_{i|0}-E(Y_{i,0}|X_i=0))\\
\label{CC}&\quad+(1-\pi_i)(r_{i|1}-E(Y_{i,1}|X_i=0))\\
\label{DD}&\quad-\pi_i(r_{i|0}-E(Y_{i,0}|X_i=1)).
\end{align}
\end{subequations}
We label the expression of (\ref{AA}) and (\ref{BB}) as
\begin{align}
   \nonumber \delta_{i,1}:=&~\pi_i(r_{i|1}-E(Y_{i,1}|X_i=1))\\
\label{AAA}&-(1-\pi_i)(r_{i|0}-E(Y_{i,0}|X_i=0))
\end{align}
and the expression of (\ref{CC}) and (\ref{DD}) as
\begin{align}
   \nonumber \delta_{i,2}:=&~(1-\pi_i)(r_{i|1}-E(Y_{i,1}|X_i=0))\\
\label{BBB}&-\pi_i(r_{i|0}-E(Y_{i,0}|X_i=1)),
\end{align}
and note that $K_i=\delta_{i,1}+\delta_{i,2}$. 

The weighted difference $\delta_{i,1}$ in (\ref{AAA}) is a measure of individual, inconsistency bias. The weighted difference $\delta_{i,2}$ in (\ref{BBB}) is a measure of an expected effect of versions on prognosis probability for the outcome.
\section{Applications}
\label{apps}
In this section, we describe three example applications in the following contexts: golf, drug use, and vaccination. Intuition is developed in the golf context. In the drug context, $\delta_{i,1}\neq 0$, and in the vaccination context, $\delta_{i,2}\neq 0$.
\subsection{Does putting cause birdies?}
\label{bird}
Each of $N$ golfers is observed playing on a par 3 hole. Treatment is putting on the second shot, and a successful outcome is a birdie (or a hole in one). For each $i$, we have recorded $x_i=1$ for the treatment and $x_i=0$ otherwise, and $y_i=1$ for a successful outcome and $y_i=0$ otherwise. 

Tjur's coefficient of discrimination measures the proportion of variation of the observed $\{x_i\}_{i=1}^N$ values explained by individual propensity probabilities $\{\pi_i\}_{i=1}^N$ \cite[Theorem 3.3]{KnaEnt}. Then, again, a separate Tjur's coefficient of discrimination measures the proportion of variation of the observed $\{y_i\}_{i=1}^N$ values explained by individual, expected prognosis probabilities $\{r_i\}_{i=1}^N$ \cite[Corollary 3.4]{KnaEnt}. Here, for the sake of argument, we assume that both of those coefficients of discrimination are $50\%$. The reader may want to imagine an uneven putting surface or windy playing conditions to help make those assumptions plausible. 

With those specified coefficients of discrimination and the data of Table \ref{tab1}, we determine $f=0.125$ and $g\approx 0.03$, and then solve (\ref{probs}) to obtain $0.05\leq \psi\leq 0.43$. We conclude that background confounding can not fully explain away the observed association.

Can background confounding combined with version confounding fully explain away the observed association? A golfer may play differently under an intervention to force the play of a certain club on the second shot, but many golfers, on many par 3 holes, aim as close to the pin as possible on the first shot. We therefore reason that the distribution of versions after the first shot is similar regardless of whether the golfer is playing naturally or under either forced intervention, and we accept (\ref{first}). 

We may reason about $K$ by focusing on the primary version, which in this context is the position of the ball after the first shot. Some positions are extremely close to the hole, while other positions are extremely far from the hole. Other positions are moderate, on the putting green, near the hole. Conditional on position, we reason about the effect of treatment on prognosis for birdie. We reason that putting increases the chance of birdie but only from moderate distances on the green, while chipping barely increases the chance of birdie from off the green. It is that asymmetry that allows us to conclude that the expected gain $r_{i|1}-r_i(1)$ is greater than the expected gain $r_{i|0}-r_i(0)$. Since this reasoning carries over to the whole population, we conclude that $K\geq 0.05$ is plausible. The versions that lead to putting may have been what was causing the birdies, not the putting itself.

Alternatively, we may reason about the expected effect of versions and reach the same conclusion. Under \ref{first}, we assess $\delta_{i,2}$ of (\ref{BBB}). Over a space of imagined versions, the difference $r_{i|1}-E(Y_{i,1}|X_i=0)$ is estimated to be positive and large by subtracting the chance for birdie when putting, given a version that naturally leads to chipping, from the chance for birdie when putting naturally. Likewise, the difference $r_{i|0}-E(Y_{i,0}|X_i=1)$ is estimated to be slightly negative by subtracting the chance for birdie when chipping (hypothetically), given a version that naturally leads to putting, from the chance for birdie when chipping naturally. Also, for each $i\in I$ the quantity $\delta_{i,1}$ of (\ref{AAA}) is estimated to be near zero. Since for most golfers $\delta_{i,1}\approx 0$ and $\delta_{i,2}>0.05$, we therefore conclude over the whole population of golfers that $K\geq 0.05$ is plausible.
\subsection{Does marijuana cause hard drug use?}
\label{drugsec}
The observational data shown in Table \ref{Drugs2} comes from the Population Assessment of tobacco and Health ({PATH}) study \cite{PATH}. That data was obtained without weighting. We assume non-informative censoring, and assume for each individual that the decision to use marijuana is made prior to the decision to use hard drugs. The category of hard drug use includes cocaine, methamphetamine, and heroin. There is a strong association between marijuana use and hard drug use. The relative risk is $11.4$. 

\begin{table}[ht!]
\centering
\caption{A contingency table of frequencies showing a positive association between marijuana use and hard drug use ($N=6,605$)}
\label{Drugs2}
\begin{tabular}{rcc}
\toprule
 & \multicolumn{2}{c}{Marijuana use}\\
\cmidrule{2-3}
Hard drug use&No&Yes\\
\hline
Yes&114&978\\
No&3,649&1,864\\
\hline
\end{tabular}
\end{table}

Measures of probandwise concordance from studies on identical twins have been reported to be around $50\%$ for marijuana use and around $40\%$ for hard drug use \cite{T96,Kendler1998,Kendler1998b}. Discordance in studies of identical twins is evidence for randomness in the data generating process, and it leads to values for the bounds $f=0.03$ and $g=0.04$; see (\ref{probs}) and \cite{knaeblePartial}. That randomness plays out starting at conception, meaning that each individual trial run starts at conception. With those parameters in place, we solve the optimization problems in (\ref{probs}) to bound $\psi$ as follows:
\begin{equation}\label{pi}0<L=0.15\leq\psi\leq 0.52=U.\end{equation}

We interpret (\ref{pi}) as evidence for the existence of chance events after conception that cause hard drug use. It is not possible for genomic variables or environmental state variables, defined at the times of conception, to fully confound the association in Table \ref{Drugs2}. However, it is not yet clear whether the primary cause of hard drug use is marijuana use or some other event or set of events occurring sometime after conception as a result of $\{\omega_i\in\Omega_i\}_{i\in I}$. There could be hidden versions of treatment that affect the outcome, e.g. alcohol use or a series of unfortunate events. Moreover, we may imagine lives lived under an interventional process with marijuana prohibited and compare that process to the more natural processes that produced the data in Table \ref{Drugs2}. All things considered, it appears that $K\geq 0.15$ is plausible. There is some evidence for the existence of one or more post-conception events that cause hard drug use, but we can't be certain that marijuana use is such an event.
\subsection{On the effectiveness of vaccines}
\label{vac}
During the Covid-19 pandemic, some people observed and reported that Covid-19 cases increased after vaccination \cite{Morris2024}. Such an increase occurred in Singapore, but in Singapore there was a strict lockdown in place for 18 months until just prior to when mass vaccination began \cite{Morris2024}. The ending of the lockdown has been shown to be a relevant co-intervention \cite{Morris2024}, and here we may think of it as one of many relevant versions of mass vaccination at the city-state level. It is easy to imagine many versions of treatment at the city-state level, but important to remember that versions of treatment exist also on the individual level.

There was a randomized trial that studied the effectiveness of the BNT162b2 mRNA Covid-19 Vaccine \cite{Polack2020} on the individual (personal) level. 
Data from that trial are shown in Table \ref{Vaccines}. The proportion of trial participants who developed Covid-19 was $0.7\%$ in the unvaccinated group and $0.04\%$ in the vaccinated group. The relative risk was $0.05$, which is equivalent to a claim that the vaccine was $95\%$ effective. The risk difference was $-0.64\%$.

\begin{table}[ht!]
\centering
\caption{A contingency table of frequencies arising from a randomized trial showing a protective association between vaccination and Covid-19 cases ($N=43,448$).}
\label{Vaccines}
\begin{tabular}{rcc}
\toprule
 & \multicolumn{2}{c}{Vaccination}\\
\cmidrule{2-3}
Case&No&Yes\\
\hline
Yes&162&8\\
No&21,558&21,720\\
\hline
\end{tabular}
\end{table}

The randomization ensures that $L=-0.64\%=\psi=-0.64\%=U$, i.e. we can identify $\psi$ with the observed, population risk difference. However, randomization alone does not justify identification of $\tau$ and $\psi$, but since the study was placebo-controlled and double-blinded \cite{Polack2020}, it is reasonable to assume $K\approx 0$, which implies $\tau \approx \psi =-0.64\%$. Hypothetically, with unblinded observers, a $K$ of slightly larger magnitude is plausible, and with unblinded participants, $K<-0.64\%$ is plausible. 

In fact, in the open-label phase of the same trial, effectiveness was no longer apparent \cite[Table 1]{Baden2024}, while safety was similar to that previously reported for the blinded part of the study \cite{Baden2024}. 
Those results are consistent with insights gained from the $K$ parameter and economic knowledge about how people optimize \cite[p. 18]{Cunningham2021}. Participants of the open-label trial \cite{Baden2024} who were assigned to placebo may have washed their hands more in an attempt to avoid Covid-19 (like the co-interventions innovated by AIDS patients \cite[p. 12]{Imbens2015}), while those assigned to the vaccine may wash their hands less due to a belief in the effectiveness of the vaccine (like what happened in Singapore). For many individuals, bias toward the null, namely \begin{equation}\label{negvers}r_i(0)>r_{i|0}=r_{i|1}>r_i(1),\end{equation} is thus a possibility, but perhaps only for the prominent Covid-19 outcome. A $K$ value of smaller magnitude is likely for an obscure safety outcome.

To better understand (\ref{negvers}), we note that (\ref{first}) is reasonable in this context, and we clarify that the treatment is vaccination only, not vaccination and awareness of vaccination. We therefore reason that $r_{i|1}>E(Y_{i,1}|X_i=0)$ because $r_{i|1}$ involves vaccination and an awareness of vaccination while $E(Y_{i,1}|X_i=0)$ involves vaccination under a belief that no vaccine was given and a precautionary state of mind. Likewise, we reason that $r_{i|0}<E(Y_{i,0}|X_i=1)$ because $r_{i|0}$ involves presumably effective \cite[p. 18]{Cunningham2021} co-interventions while $E(Y_{i,0}|X_i=1)$ may involve some risky behavior due to belief in a protective vaccine.
\section{Discussion}
The main problem is the following: that which was (quasi) randomly assigned may not be precisely the treatment. Nature might naturally ``assign'' more than just the treatment, and 
versions of treatment may not be balanced. 
Here, we are not discussing chance imbalance \cite{Morgan2012} but rather version confounding (see \ref{longconf}), which is more systematic and ultimately a problem of study design. 

The problem of hidden versions of treatment can be addressed theoretically by redefining the treatment \cite[p. 11]{Imbens2015}. However, there may be predefined treatments and consistency violations of practical concern \cite{Rehkopf2016}. Consistency has been viewed as an axiom, a definition, a theorem, and an assumption \cite{Pearl2010b}, and while consistency may not be an assumption in the conventional sense \cite{RosePrivate}, it can be formulated as an assumption \cite{Cole2009,VanderWeele2009} and recognized fundamentally as an exclusion restriction \cite[p. 12]{Imbens2015}. 

To avoid assumptions and address practical problems related to consistency violations, this paper has introduced a two-staged methodology for sensitivity analysis with stochastic potential outcomes. It supports causal inference from observational studies and randomized experiments. 

The technical methodology of this paper amounts to a mapping (or algorithm), given the observed data $\{x_i,y_i\}_{i\in I}$. The inputs to the mapping are two entropy parameters, $f$ and $g$, and one bias parameter, $K$. The output is an interval that partially identifies the causal estimand, $\tau$.

The introduced methodology is an alternative to instrumental variables analysis and intention to treat analysis. The methodology does not require Laplacian determinism \cite{Dawid2012} or a definition of ($\mathcal{F}_{0,1}$-measurable) compliers. Even with some noncompliance, entropy in the data generating process lowers the $f$ and $g$ parameters of (\ref{probs}), thus tightening the bounds of partial identification on $\psi$. But, unlike intention to treat analysis where treatment and control groups are defined by assignment, here treatment and control groups are defined by whether or not the treatment is actually taken, and the parameter $K$ accounts for versions of treatment that arise along the way.
%
%
%
%
%
\section*{Acknowledgements}
The authors thank Andrew Gelman, Sam Pimentel, and Machiel van Frankenhuijsen for helping to craft the main ideas behind this manuscript. The authors thank Tom Alberts for valuable feedback. 
\bibliography{refs}

@article{angrist1994ident,
  title={Identification and Estimation of Local Average Treatment Effects},
  journal={Econometrica},
  volume={62},
  number={2},
  pages={467--475},
  year={1994},
  author={Angrist, Joshua D and Imbens, Guido W}
}

@article{angrist1990lifetime,
  title={Lifetime Earnings and the Vietnam Era Draft Lottery: Evidence from Social Security Administrative Records},
  author={Angrist, Joshua D},
  journal={The American Economic Review},
  volume={80},
  number={5},
  pages={1284-1286},
  year={1990}
}

@article{Baden2024,
  title = {Long-term safety and effectiveness of mRNA-1273 vaccine in adults: COVE trial open-label and booster phases},
  volume = {15},
  ISSN = {2041-1723},
  url = {http://dx.doi.org/10.1038/s41467-024-50376-z},
  DOI = {10.1038/s41467-024-50376-z},
  number = {1},
  journal = {Nature Communications},
  publisher = {Springer Science and Business Media LLC},
  author = {Baden,  Lindsey R. and El Sahly,  Hana M. and Essink,  Brandon and Follmann,  Dean and Hachigian,  Gregory and Strout,  Cynthia and Overcash,  J. Scott and Doblecki-Lewis,  Susanne and Whitaker,  Jennifer A. and Anderson,  Evan J. and Neuzil,  Kathleen and Corey,  Lawrence and Priddy,  Frances and Tomassini,  Joanne E. and Brown,  Mollie and Girard,  Bethany and Stolman,  Dina and Urdaneta,  Veronica and Wang,  Xiaowei and Deng,  Weiping and Zhou,  Honghong and Dixit,  Avika and Das,  Rituparna and Miller,  Jacqueline M.},
  year = {2024},
  month = aug 
}

@article{Baiocchi2014,
  title = {Instrumental variable methods for causal inference},
  volume = {33},
  ISSN = {0277-6715},
  url = {http://dx.doi.org/10.1002/sim.6128},
  DOI = {10.1002/sim.6128},
  number = {13},
  journal = {Statistics in Medicine},
  publisher = {Wiley},
  author = {Baiocchi,  Michael and Cheng,  Jing and Small,  Dylan S.},
  year = {2014},
  month = mar,
  pages = {2297–2340}
}

@book{Billing,
  author = {Billingsley, {Patrick}},
  edition = {Third Edition},
  isbn = {978-81-265-1771-8},
  publisher = {Wiley},
  title = {Probability and measure},
  year = {1995/2017}
}

@article{Cole2009,
  title = {The Consistency Statement in Causal Inference: A Definition or an Assumption?},
  volume = {20},
  ISSN = {1044-3983},
  url = {http://dx.doi.org/10.1097/EDE.0b013e31818ef366},
  DOI = {10.1097/ede.0b013e31818ef366},
  number = {1},
  journal = {Epidemiology},
  publisher = {Ovid Technologies (Wolters Kluwer Health)},
  author = {Cole,  Stephen R. and Frangakis,  Constantine E.},
  year = {2009},
  month = jan,
  pages = {3–5}
}

@book{Cunningham2021,
  title = {Causal Inference: The Mixtape},
  ISBN = {9780300251685},
  url = {http://dx.doi.org/10.2307/j.ctv1c29t27},
  DOI = {10.2307/j.ctv1c29t27},
  publisher = {Yale University Press},
  author = {Cunningham,  Scott},
  year = {2021},
  month = jan 
}

@article{Dawid2012,
  title = {“Imagine a Can Opener”--The Magic of Principal Stratum Analysis},
  volume = {8},
  ISSN = {1557-4679},
  url = {http://dx.doi.org/10.1515/1557-4679.1391},
  DOI = {10.1515/1557-4679.1391},
  number = {1},
  journal = {The International Journal of Biostatistics},
  publisher = {Walter de Gruyter GmbH},
  author = {Dawid,  Philip and Didelez,  Vanessa},
  year = {2012},
  month = jan 
}

@book{deming1994new,
  title={The New Economics for Industry, Government, Education},
  author={Deming, W. Edwards},
  year={1994},
  edition={2nd},
  publisher={MIT Center for Advanced Engineering Study},
  address={Cambridge, MA}
}

@article{SAWA,
	doi = {10.1097/ede.0000000000000457},
	url = {https://doi.org/10.1097%2Fede.0000000000000457},
	year = 2016,
	month = {may},
	publisher = {Ovid Technologies (Wolters Kluwer Health)},
	volume = {27},
	number = {3},
	pages = {368--377},
	author = {Peng Ding and Tyler J. VanderWeele},
	title = {Sensitivity Analysis Without Assumptions},
	journal = {Epidemiology}
}

@article{Frangakis2002,
  title = {Principal Stratification in Causal Inference},
  volume = {58},
  ISSN = {1541-0420},
  url = {http://dx.doi.org/10.1111/j.0006-341X.2002.00021.x},
  DOI = {10.1111/j.0006-341x.2002.00021.x},
  number = {1},
  journal = {Biometrics},
  publisher = {Oxford University Press (OUP)},
  author = {Frangakis,  Constantine E. and Rubin,  Donald B.},
  year = {2002},
  month = mar,
  pages = {21–29}
}

@article{Gelman2025,
  title = {Russian roulette: The need for stochastic potential outcomes when utilities depend on counterfactuals},
  ISSN = {1464-3510},
  url = {http://dx.doi.org/10.1093/biomet/asaf062},
  DOI = {10.1093/biomet/asaf062},
  journal = {Biometrika},
  publisher = {Oxford University Press (OUP)},
  author = {Gelman,  Andrew and Mikhaeil,  Jonas M},
  year = {2025},
  month = aug 
}

@article{Hasegawa2020,
  title = {Causal Inference With Two Versions of Treatment},
  volume = {45},
  ISSN = {1935-1054},
  url = {http://dx.doi.org/10.3102/1076998620914003},
  DOI = {10.3102/1076998620914003},
  number = {4},
  journal = {Journal of Educational and Behavioral Statistics},
  publisher = {American Educational Research Association (AERA)},
  author = {Hasegawa,  Raiden B. and Deshpande,  Sameer K. and Small,  Dylan S. and Rosenbaum,  Paul R.},
  year = {2020},
  month = mar,
  pages = {426–445}
}

@book{Imbens2015,
  doi = {10.1017/cbo9781139025751},
  url = {https://doi.org/10.1017/cbo9781139025751},
  year = {2015},
  month = apr,
  publisher = {Cambridge University Press},
  author = {Guido W. Imbens and Donald B. Rubin},
  title = {Causal Inference for Statistics,  Social,  and Biomedical Sciences}
}

@article{Kendler1998,
  doi = {10.1176/ajp.155.8.1016},
  url = {https://doi.org/10.1176/ajp.155.8.1016},
  year = {1998},
  month = aug,
  publisher = {American Psychiatric Association Publishing},
  volume = {155},
  number = {8},
  pages = {1016--1022},
  author = {Kenneth S. Kendler and Carol A. Prescott},
  title = {Cannabis Use,  Abuse,  and Dependence in a Population-Based Sample of Female Twins},
  journal = {American Journal of Psychiatry}
}

@article{Kendler1998b,
  doi = {10.1192/bjp.173.4.345},
  url = {https://doi.org/10.1192/bjp.173.4.345},
  year = {1998},
  month = oct,
  publisher = {Royal College of Psychiatrists},
  volume = {173},
  number = {4},
  pages = {345--350},
  author = {Kenneth S. Kendler and Carol A. Prescott},
  title = {Cocaine use,  abuse and dependence in a population-based sample of female twins},
  journal = {British Journal of Psychiatry}
}

@misc{KnaEnt,
  doi = {10.48550/ARXIV.2407.08862},
  url = {https://arxiv.org/abs/2407.08862},
  author = {Knaeble,  Brian and Hakim-Hashemi,  Mehdi and Abramson,  Mark A.},
  title = {Maximum Entropy Estimation of Heterogeneous Causal Effects},
  publisher = {arXiv},
  year = {2024},
  copyright = {Creative Commons Attribution 4.0 International}
}

@article{Knaeble2023,
  doi = {10.1002/sta4.609},
  url = {https://doi.org/10.1002/sta4.609},
  year = {2023},
  month = aug,
  publisher = {Wiley},
  volume = {12},
  number = {1},
  author = {Brian Knaeble and Braxton Osting and Placede Tshiaba},
  title = {An asymptotic threshold of sufficient randomness for causal inference},
  journal = {Stat}
}

@article{Knaeble2024,
  title = {Observational Causality Testing},
  volume = {13},
  ISSN = {2049-1573},
  url = {http://dx.doi.org/10.1002/sta4.70017},
  DOI = {10.1002/sta4.70017},
  number = {4},
  journal = {Stat},
  publisher = {Wiley},
  author = {Knaeble,  Brian and Osting,  Braxton and Tshiaba,  Placede},
  year = {2024},
  month = oct 
}

@misc{knaeblePartial,
      title={Partial Identification of the Average Treatment Effect with Stochastic Counterfactuals and Discordant Twins}, 
      author={Brian Knaeble and Braxton Osting and Placede Tshiaba},
      year={2024},
      eprint={2407.19057},
      archivePrefix={arXiv},
      primaryClass={stat.ME},
      url={https://arxiv.org/abs/2407.19057}, 
}

@article{Mellon2020,
  title = {Rain,  Rain,  Go Away: 137 Potential Exclusion-Restriction Violations for Studies Using Weather as an Instrumental Variable},
  ISSN = {1556-5068},
  url = {http://dx.doi.org/10.2139/ssrn.3715610},
  DOI = {10.2139/ssrn.3715610},
  journal = {SSRN Electronic Journal},
  publisher = {Elsevier BV},
  author = {Mellon,  Jonathan},
  year = {2020}
}

@article{Morgan2012,
  title = {Rerandomization to improve covariate balance in experiments},
  volume = {40},
  ISSN = {0090-5364},
  url = {http://dx.doi.org/10.1214/12-AOS1008},
  DOI = {10.1214/12-aos1008},
  number = {2},
  journal = {The Annals of Statistics},
  publisher = {Institute of Mathematical Statistics},
  author = {Morgan,  Kari Lock and Rubin,  Donald B.},
  year = {2012},
  month = apr 
}

@article{Morris2024,
  title = {Tracking vaccine effectiveness in an evolving pandemic,  countering misleading hot takes and epidemiologic fallacies},
  volume = {194},
  ISSN = {1476-6256},
  url = {http://dx.doi.org/10.1093/aje/kwae280},
  DOI = {10.1093/aje/kwae280},
  number = {4},
  journal = {American Journal of Epidemiology},
  publisher = {Oxford University Press (OUP)},
  author = {Morris,  Jeffrey S},
  year = {2024},
  month = aug,
  pages = {898–907}
}

@article{Pearl2010b,
  title = {On the Consistency Rule in Causal Inference: Axiom,  Definition,  Assumption,  or Theorem?},
  volume = {21},
  ISSN = {1044-3983},
  url = {http://dx.doi.org/10.1097/EDE.0b013e3181f5d3fd},
  DOI = {10.1097/ede.0b013e3181f5d3fd},
  number = {6},
  journal = {Epidemiology},
  publisher = {Ovid Technologies (Wolters Kluwer Health)},
  author = {Pearl,  Judea},
  year = {2010},
  month = nov,
  pages = {872–875}
}

@article{Pim,
  doi = {10.1080/01621459.2015.1076342},
  url = {https://doi.org/10.1080/01621459.2015.1076342},
  year = {2016},
  month = jul,
  publisher = {Informa {UK} Limited},
  volume = {111},
  number = {515},
  pages = {1157--1167},
  author = {Samuel D. Pimentel and Dylan S. Small and Paul R. Rosenbaum},
  title = {Constructed Second Control Groups and Attenuation of Unmeasured Biases},
  journal = {Journal of the American Statistical Association}
}

@ARTICLE{Polack2020,
  title     = "Safety and efficacy of the {BNT162b2} {mRNA} Covid-19 vaccine",
  author    = "Polack, Fernando P and Thomas, Stephen J and Kitchin, Nicholas
               and Absalon, Judith and Gurtman, Alejandra and Lockhart, Stephen
               and Perez, John L and P{\'e}rez Marc, Gonzalo and Moreira, Edson
               D and Zerbini, Cristiano and Bailey, Ruth and Swanson, Kena A
               and Roychoudhury, Satrajit and Koury, Kenneth and Li, Ping and
               Kalina, Warren V and Cooper, David and Frenck, Jr, Robert W and
               Hammitt, Laura L and T{\"u}reci, {\"O}zlem and Nell, Haylene and
               Schaefer, Axel and {\"U}nal, Serhat and Tresnan, Dina B and
               Mather, Susan and Dormitzer, Philip R and {\c S}ahin, U{\u g}ur
               and Jansen, Kathrin U and Gruber, William C and {C4591001
               Clinical Trial Group}",
  journal   = "N. Engl. J. Med.",
  publisher = "Massachusetts Medical Society",
  volume    =  383,
  number    =  27,
  pages     = "2603--2615",
  month     =  dec,
  year      =  2020,
  copyright = "http://www.nejmgroup.org/legal/terms-of-use.htm",
  language  = "en"
}

@article{Rehkopf2016,
  title = {The Consistency Assumption for Causal Inference in Social Epidemiology: When a Rose Is Not a Rose},
  volume = {3},
  ISSN = {2196-2995},
  url = {http://dx.doi.org/10.1007/s40471-016-0069-5},
  DOI = {10.1007/s40471-016-0069-5},
  number = {1},
  journal = {Current Epidemiology Reports},
  publisher = {Springer Science and Business Media LLC},
  author = {Rehkopf,  David H. and Glymour,  M. Maria and Osypuk,  Theresa L.},
  year = {2016},
  month = feb,
  pages = {63–71}
}

@book{Rose10,
  title={Design of Observational Studies},
  author={Rosenbaum, P.R.},
  isbn={9781441912138},
  lccn={2009938109},
  series={Springer Series in Statistics},
  url={https://books.google.com/books?id=hRtHAAAAQBAJ},
  year={2009},
  publisher={Springer New York}
}

@article{Rosen,
  doi = {10.1257/jel.38.4.827},
  url = {https://doi.org/10.1257/jel.38.4.827},
  year = {2000},
  month = dec,
  publisher = {American Economic Association},
  volume = {38},
  number = {4},
  pages = {827--874},
  author = {Mark R Rosenzweig and Kenneth I Wolpin},
  title = {Natural {\textquotedblleft}Natural Experiments{\textquotedblright} in Economics},
  journal = {Journal of Economic Literature}
}

@ARTICLE{Rubin2009,
  title     = "Should observational studies be designed to allow lack of
               balance in covariate distributions across treatment groups?",
  author    = "Rubin, Donald B",
  journal   = "Stat. Med.",
  publisher = "Wiley",
  volume    =  28,
  number    =  9,
  pages     = "1420--1423",
  month     =  apr,
  year      =  2009,
  copyright = "http://onlinelibrary.wiley.com/termsAndConditions\#vor",
  language  = "en"
}

@article{Tjur2009,
  title = {Coefficients of Determination in Logistic Regression Models—A New Proposal: The Coefficient of Discrimination},
  volume = {63},
  ISSN = {1537-2731},
  url = {http://dx.doi.org/10.1198/tast.2009.08210},
  DOI = {10.1198/tast.2009.08210},
  number = {4},
  journal = {The American Statistician},
  publisher = {Informa UK Limited},
  author = {Tjur,  Tue},
  year = {2009},
  month = nov,
  pages = {366–372}
}

@MISC{PATH,
  title     = "Population Assessment of tobacco and Health ({PATH}) study
               [United States] public-use files",
  author    = "{US NIH et al.}",
  publisher = "ICPSR - Interuniversity Consortium for Political and Social
               Research",
  year      =  2016
}

@misc{RosePrivate,
  author = "Paul Roansenbaum",
  howpublished = "private communication",
  year = "2025"
}

@article{T96,
  doi = {10.1002/(sici)1096-8628(19960920)67:5<473::aid-ajmg6>3.0.co;2-l},
  url = {https://doi.org/10.1002/(sici)1096-8628(19960920)67:5<473::aid-ajmg6>3.0.co;2-l},
  year = {1996},
  month = sep,
  publisher = {Wiley},
  volume = {67},
  number = {5},
  pages = {473--477},
  author = {Ming T. Tsuang and Michael J. Lyons and Seth A. Eisen and Jack Goldberg and William True and Nong Lin and Joanne M. Meyer and Rosemary Toomey and Stephen V. Faraone and Lindon Eaves},
  title = {Genetic influences {onDSM}-{III}-R drug abuse and dependence: A study of 3, 372 twin pairs},
  journal = {American Journal of Medical Genetics}
}

@article{VanderWeele2009,
  title = {Concerning the Consistency Assumption in Causal Inference},
  volume = {20},
  ISSN = {1044-3983},
  url = {http://dx.doi.org/10.1097/EDE.0b013e3181bd5638},
  DOI = {10.1097/ede.0b013e3181bd5638},
  number = {6},
  journal = {Epidemiology},
  publisher = {Ovid Technologies (Wolters Kluwer Health)},
  author = {VanderWeele,  Tyler J.},
  year = {2009},
  month = nov,
  pages = {880–883}
}

@article{VanderWeele2013,
  title = {Causal inference under multiple versions of treatment},
  volume = {1},
  ISSN = {2193-3677},
  url = {http://dx.doi.org/10.1515/jci-2012-0002},
  DOI = {10.1515/jci-2012-0002},
  number = {1},
  journal = {Journal of Causal Inference},
  publisher = {Walter de Gruyter GmbH},
  author = {VanderWeele,  Tyler J. and Hernan,  Miguel A.},
  year = {2013},
  month = may,
  pages = {1–20}
}

@article{VanderWeele2018,
  doi = {10.1097/ede.0000000000000823},
  url = {https://doi.org/10.1097/ede.0000000000000823},
  year = {2018},
  month = jul,
  publisher = {Ovid Technologies (Wolters Kluwer Health)},
  volume = {29},
  number = {4},
  pages = {e24--e25},
  author = {Tyler J. VanderWeele},
  title = {On Well-defined Hypothetical Interventions in the Potential Outcomes Framework},
  journal = {Epidemiology}
}

@article{Weichenthal2022,
  title = {Fine Particulate Air Pollution and the “No-Multiple-Versions-of-Treatment” Assumption: Does Particle Composition Matter for Causal Inference?},
  volume = {192},
  ISSN = {1476-6256},
  url = {http://dx.doi.org/10.1093/aje/kwac191},
  DOI = {10.1093/aje/kwac191},
  number = {2},
  journal = {American Journal of Epidemiology},
  publisher = {Oxford University Press (OUP)},
  author = {Weichenthal,  Scott and Ripley,  Susannah and Korsiak,  Jill},
  year = {2022},
  month = nov,
  pages = {147–153}
}

@article{Wool,
  doi = {10.1016/j.rie.2016.01.001},
  url = {https://doi.org/10.1016/j.rie.2016.01.001},
  year = {2016},
  month = jun,
  publisher = {Elsevier {BV}},
  volume = {70},
  number = {2},
  pages = {232--237},
  author = {Jeffrey M. Wooldridge},
  title = {Should instrumental variables be used as matching variables?},
  journal = {Research in Economics}
}

@misc{Deanamuir,
  author = {Deanamuir},
  year = {2020},
  title = {picturesque par 3 9th at The Machrie Links},
  note = {Retreived on 11/12/2025 from Wikimedia Commons at \url{https://commons.wikimedia.org/wiki/File:9th_bunkers_(2).jpg}. Licensed under CC BY-SA 4.0, \url{https://creativecommons.org/licenses/by-sa/4.0/}}
}
\bibliographystyle{plain}
\end{document}